\documentclass{article}

\usepackage{amssymb}

\newcommand{\Tr}{\mathop{\mathrm{Tr}}}

\newcommand{\bbbone}{{\mathchoice {\rm 1\mskip-4mu l} {\rm 1\mskip-4mu l}
{\rm 1\mskip-4.5mu l} {\rm 1\mskip-5mu l}}}

\newcommand{\ui}{\mathrm{i}}
\newcommand{\dd}{\mathrm{d}}

\newtheorem{theorem}{Theorem}

\begin{document}

\title{Entropy and information gain  in \\ quantum continual measurements}

\author{Alberto Barchielli \\
\small{Politecnico di Milano, Dipartimento di Matematica,} \\ \small{Piazza
Leonardo da Vinci 32, I-20133 Milano, Italy;} \\ \small{also Istituto Nazionale
di Fisica Nucleare, Sezione di Milano.} \\ \small{E-mail:
barchielli@mate.polimi.it}}
\date{}
\maketitle

\section{Introduction}\label{S1}

The theory of measurements continuous in time in quantum mechanics (quantum
continual measurements) has been formulated by using the notions of instrument,
positive operator valued (POV) measure, etc.\ \cite{Dav76,BarLP83}, by using
quantum stochastic differential equations \cite{BarL85,Bar90} and by using
classical stochastic differential equations (SDE's) for vectors in Hilbert
spaces or for trace-class operators \cite{Bel88,Bel89a,Bel89b,BarB91}. In the
same times Ozawa made developments in the theory of instruments
\cite{Oza84,Oza95} and introduced the related notions of \emph{a posteriori}
states \cite{Oza85} and of information gain \cite{Oza86}.

In Section \ref{S2} we introduce a simple class of SDE's relevant to the theory
of continual measurements and we recall how they are related to instruments and
\emph{a posteriori} states and, so, to the general formulation of quantum
mechanics \cite{Kra80}. In Section \ref{S3} we shall introduce and  use the
notion of information gain and the other results of paper \cite{Oza86} inside
the theory of continual measurements.

\section{Stochastic
differential equations \\ and instruments}\label{S2}

Let $\mathcal{H}$ be a separable complex Hilbert space, associated to the
quantum system of interest. Let us denote by $\mathcal{B}(\mathcal{H})$ the
space of bounded linear operators on $\mathcal{H}$ and by
$\mathcal{T}(\mathcal{H})$ the trace-class on $\mathcal{H}$, i.e.
$\mathcal{T}(\mathcal{H}) = \big\{\rho \in \mathcal{B}(\mathcal{H}): \|\rho\|
\equiv \Tr\left\{\sqrt{\rho^* \rho}\right\} < \infty \big\}$. Let
$\mathcal{S}(\mathcal{H})\subset \mathcal{T}(\mathcal{H})$ be the set of all
statistical operators (\emph{states}) on $\mathcal{H}$. Commutators and
anticommutators are denoted by $[\ , \ ]$ and $\{\ , \ \}$, respectively.

Let $H$, $L_j$, $S_h$, $ j,h=1,2,\ldots$, be bounded operators on $\mathcal{H}$
such that $H=H^\dagger$, $\sum_{j=1}^\infty L^\dagger_jL_j$ and
$\sum_{h=1}^\infty S_h^\dagger S_h$ are strongly convergent in
$\mathcal{B}(\mathcal{H})$. Let $J_k$ be a bounded linear map on
$\mathcal{T}(\mathcal{H})$ such that its adjoint $J_k^*$ is a normal,
completely positive map on $\mathcal{B}(\mathcal{H})$ and $\sum_{k=1}^\infty
J^*_k[\bbbone]$ is strongly convergent to a bounded operator. Then, we
introduce the following operators on $\mathcal{T}(\mathcal{H})$:
\begin{eqnarray}
\mathcal{L}_0[\rho] &=& - \ui [H, \rho] + \sum_{j=1}^\infty \Big(L_j \rho
L_j^\dagger - \frac{1}{2} \left\{L_j^\dagger L_j ,\,\rho\right\}\Big) \nonumber
\\
&{}&+ \sum_{k = 1}^\infty \Big(J_k[\rho] - \frac{1}{2} \left\{
J_k^*[\bbbone],\,\rho\right\}\Big), \label{2+1}
\\
\mathcal{L}_1[\rho] &=& \sum_{h=1}^\infty \Big(S_h \rho S_h^\dagger -
\frac{1}{2} \left\{S_h^\dagger S_h,\,\rho\right\}\Big), \label{2+2}
\\
\mathcal{L} &=& \mathcal{L}_0 + \mathcal{L}_1. \label{2+3}
\end{eqnarray}
The adjoint operators of $\mathcal{L}$, $\mathcal{L}_0$, $\mathcal{L}_1$ are
generators of norm-continuous quantum dynamical semigroups \cite{Lin76,AliL87}.

Let us now consider the following linear SDE (in the sense of It\^o) for
trace-class operators:
\begin{eqnarray}
\dd \sigma_t &=& \mathcal{L}[\sigma_{t^-}]\, \dd t + \sum_{j=1}^\infty
\left(\widetilde L_j(t) \sigma_{t^-} + \sigma_{t^-} \widetilde L_j(t)^\dagger
\right) \dd \widetilde W_j(t) + {} \nonumber
\\  &{}& +
\sum_{k=1}^\infty \left(\frac{1}{\lambda_k}\, J_k[\sigma_{t^-}]  - \sigma_{t^-}
\right) \bigl( \dd N_k(t)- \lambda_k\, \dd  t \bigr); \label{2+4}
\end{eqnarray}
the initial condition is $\sigma_0=\rho\in \mathcal{S}(\mathcal{H})$ (a
non-random state) and we have set
\begin{equation} \label{2+5}
\widetilde L_j(t) = {\rm e}^{\ui \omega_j t} L_j\, , \qquad \qquad \omega_j \in
\mathbb{R}\,.
\end{equation}
The processes $\widetilde W_j(t)$ are independent standard Wiener processes,
the $N_k(t)$ are independent Poisson processes of intensity $\lambda_k>0$,
which are also independent of the Wiener processes; we assume $\sum_k
\lambda_k<+\infty$.

These processes are realized in a probability space $(\Omega, \mathcal{F}, Q)$;
the sample space $\Omega$ is, roughly speaking, the set of possible
trajectories for the processes $\widetilde W_j$, $N_k$, the event space
$\mathcal{F}$ is the $\sigma$-algebra of sets of trajectories to which a
probability can be given and $Q$ is the probability law under which $\widetilde
W_j$, $N_k$ are independent Wiener and Poisson processes. Moreover, let
$\mathcal{F}_t$ be the collection of events which are specified by giving
conditions involving times only in the interval $[0,t]$. We also ask
$\mathcal{F}=\mathcal{F}_\infty$. In mathematical terms the $\widetilde W_j$,
$N_k$ are canonical Wiener and Poisson processes, $\{\mathcal{F}_t,\,t\geq 0\}$
is their natural filtration and $\mathcal{F}=\bigvee_{t\geq 0} \mathcal{F}_t$.
Finally, let us denote by $\mathbb{E}_Q$ the expectation with respect to the
probability $Q$, i.e. $\mathbb{E}_Q[A]= \int_\Omega A(\omega) Q(\dd \omega)$.

For every $F\in \mathcal{F}_t$ and every initial condition
$\rho\in\mathcal{S}(\mathcal{H})$, let us set
\begin{equation}\label{2+8}
\mathcal{I}_t(F)[\rho]= \mathbb{E}_Q[1_F \sigma_t] \equiv \int_F
\sigma_t(\omega) Q(\dd \omega);
\end{equation}
$1_F$ is the indicator function of the set $F$, i.e. $1_F(\omega)=1$ if
$\omega\in F$ and $1_F(\omega)=0$ if $\omega\notin F$. The map $\mathcal{I}_t$
turns out to be a (completely positive) instrument \cite{Oza84} with value
space $(\Omega,\mathcal{F}_t)$ and $\mathcal{I}_t(\cdot)^*[\bbbone]$ is the
associated POV measure. Then we set, $\forall F\in \mathcal{F}_t$,
\begin{equation}\label{2+9}
P_\rho(F) = \Tr \left\{ \mathcal{I}_t(F)^*[\bbbone] \rho \right\} =
\mathbb{E}_Q \left[ \left\|\sigma_t\right\| 1_F\right].
\end{equation}
The important point in this formula is that $ \left\|\sigma_t\right\| $ is a
$Q$-martingale and this implies that the time dependent probability measures on
the r.h.s.\ are consistent and define a unique probability $P_\rho$ on
$(\Omega,\mathcal{F})$.

The interpretation of eqs.\ (\ref{2+8}) and (\ref{2+9}) is that
$\{\mathcal{I}_t,\, t\geq 0\}$ is the family of instruments describing the
continual measurement, the processes $\widetilde W_j$, $N_k$ represent the
output of this measurement and $P_\rho$ is the physical probability law of the
output.

From eq.\ (\ref{2+8}) it follows that
\begin{equation}\label{2+6}
\eta_t = \mathcal{I}_t(\Omega)[\rho] =\mathbb{E}_Q [ \sigma_t ]
\end{equation}
is the state to be attributed to the system at time $t$ if the output of the
measurement is not taken into account or not known; it can be called the
\emph{a priori} state at time $t$. It turns out that the \emph{a priori} states
satisfy the master equation
\begin{equation}\label{2+7}
\frac{\dd\ }{\dd t}\, \eta_t = \mathcal{L}[\eta_t]\,, \qquad \eta_0=\rho\,.
\end{equation}

If we introduce the random states
\begin{equation}\label{2+10}
\rho_t= \frac{\sigma_t}{ \left\|\sigma_t\right\|}\,,
\end{equation}
then we have, $\forall F\in \mathcal{F}_t$,
\begin{equation}\label{2+11}
\mathcal{I}_t(F)[\rho] =\mathbb{E}_Q[1_F\sigma_t]=
\mathbb{E}_{P_\rho}\left[1_F\,\frac{\sigma_t}{ \left\|\sigma_t\right\|}\right]=
\int_F \rho_t(\omega) P_\rho(\dd \omega).
\end{equation}
According to \cite{Oza85}, $\rho_t(\omega)$ is a family of \emph{a posteriori}
states for the instrument $\mathcal{I}_t$ and the initial state $\rho$, i.e.\
$\rho_t(\omega)$ is the state to be attributed to the system at time $t$ when
the trajectory $\omega$ of the output is known, up to time $t$. Note that
$\eta_t = \mathbb{E}_Q [ \sigma_t ] = \mathbb{E}_{P_\rho}[\rho_t]$.

By using It\^o's calculus, we find that the \emph{a posteriori} states satisfy
the non-linear SDE
\begin{eqnarray}
\dd \rho_t &=& \mathcal{L}\left[\rho_{t^-}\right] \dd t + \sum_{j=1}^\infty
\left[\widetilde L_j(t) \rho_{t^-} + \rho_{t^-} \widetilde L_j(t)^\dagger -
m_j(t) \rho_{t^-} \right] \dd W_j(t) + {} \nonumber
\\  &{}& +
\sum_{k=1}^\infty \left[\frac{1}{\nu_k(t)}\, J_k[\rho_{t^-}]  - \rho_{t^-}
\right] \bigl( \dd N_k(t)- \nu_k(t)\, \dd  t \bigr), \label{2+12}
\end{eqnarray}
where
\begin{equation}\label{2+13}
W_j(t) = \widetilde W_j(t) - \int_0^t m_j(s)\, \dd s\, ,
\end{equation}
\begin{equation}\label{2+14}
m_j(t) = \Tr\left\{ \rho_{t^-}\left( \widetilde L_j(t) + \widetilde
L_j(t)^\dagger\right)\right\}, \qquad \nu_k(t)  = \Tr \left\{
\rho_{t^-}J^*_k[\bbbone] \right\}.
\end{equation}
Under the physical probability law $P_\rho$, the processes $W_j(t)$ are
independent standard Wiener processes and the $N_k(t)$ are counting processes
with stochastic intensity $\nu_k(t)$. In eq.\ (\ref{2+12}) the sum in the jump
term is only on the set where the stochastic intensity $\nu_k(t)$ is different
from zero.

Formulae for the moments of the output can be obtained by the technique of the
characteristic operator \cite{BarLP83,BarL85,Bar90}. Let  $h_{k\alpha}$ be real
test functions in a suitable space; we define the characteristic operator
$\mathcal{G}$ by
\begin{eqnarray}\nonumber
\mathcal{G}_t(h)[\rho]&=& \mathbb{E}_{P_\rho} \bigg[ \exp \bigg\{ \ui \sum_j
\int_0^t h_{j1}(s)\, \dd \widetilde W_j(s)
\\ &&\null\qquad {}+ \ui \sum_k \int_0^t h_{k2}(s) \,
\dd N_k(s) \bigg\}\, \rho_t \bigg]\,; \label{2+16}
\end{eqnarray}
then, $\Tr\big\{\mathcal{G}_t(h)[\rho]\big\}$ is the characteristic functional
of the output up to time $t$ (the Fourier transform of $P_\rho$ restricted to
$\mathcal{F}_t$ ). By It\^o's calculus we obtain
\begin{equation}\label{2+17}
\frac{\dd \ }{\dd t}\,\mathcal{G}_t(h)[\rho]= \mathcal{K}_t(h) \circ
\mathcal{G}_t(h)[\rho]\,,
\end{equation}
\begin{eqnarray}\nonumber
\mathcal{K}_t(h)[\rho] &=& \mathcal{L}[\rho] +\ui \sum_j h_{j1}(t) \left[
\widetilde L_j(t) \rho + \rho \widetilde L_j(t)^\dagger \right]
\\&{}&
- \frac 1 2 \sum_j h_{j1}(t)^2 \rho + \sum_k \left\{ \exp\left[\ui
h_{k2}(t)\right] - 1 \right\} J_k[\rho]\,.\label{2+18}
\end{eqnarray}
All the moments can be obtained by functional differentiation of the
characteristic functional. In particular, the mean values are expressed in
terms of the \emph{a priori} states as
\[
\mathbb{E}_{P_\rho}\left[ \widetilde W_j(t) \right] = \int_0^t
\mathbb{E}_{P_\rho}\left[ m_j(s)\right] \dd s\,, \quad
\mathbb{E}_{P_\rho}\left[ N_k(t) \right] = \int_0^t \mathbb{E}_{P_\rho}\left[
\nu_k(s)\right] \dd s\,,
\]
\[
\mathbb{E}_{P_\rho}\left[ m_j(s)\right] = \Tr\left\{\eta_s \left( \widetilde
L_j(s) + \widetilde L_j(s)^\dagger \right)\right\}\,, \quad
\mathbb{E}_{P_\rho}\left[ \nu_k(s)\right] = \Tr\left\{ J_k[\eta_s]\right\}\,,
\]
and the second moments are given by
\begin{eqnarray*}
&&\mathbb{E}_{P_\rho} \big[ X_{j\alpha}(t) X_{i\beta}(s) \big] = \delta_{ij}
\delta_{\alpha \beta} \int_0^{\min\{t,s\}} \dd \tau \left( \delta_{\alpha 1}
+\delta_{\alpha 2} \Tr \big\{ J_i[\eta_\tau]\big\} \right)\quad\null{}
\\  &{}&+
\int_0^t \dd \tau_1 \int_0^{\min\{s,\tau_1\}} \dd \tau_2 \Tr \left\{
\mathcal{A}_{j\alpha}(\tau_1) \circ \mathrm{e}^{\mathcal{L} (\tau_1-\tau_2)}
\circ \mathcal{A}_{i\beta}(\tau_2)[\eta_{\tau_2}]\right\}
\\ &{}&+
\int_0^s \dd \tau_2 \int_0^{\min\{t,\tau_2\}} \dd \tau_1 \Tr \left\{
\mathcal{A}_{i\beta}(\tau_2) \circ \mathrm{e}^{\mathcal{L} (\tau_2-\tau_1)}
\circ \mathcal{A}_{j\alpha}(\tau_1)[\eta_{\tau_1}]\right\},
\end{eqnarray*}
where $X_{j1}(t) = \widetilde W_j(t)$, $X_{j2}(t) = N_j(t)$,
$\mathcal{A}_{j1}(t)[\rho]= \widetilde L_j(t) \rho + \rho \widetilde
L_j(t)^\dagger$, $\mathcal{A}_{j2}(t)= J_j$.

The class of SDE's presented here is a particular case of the one studied in
\cite{BarH95} and, while not so general, it contains the main detection schemes
found in quantum optics \cite{BarP96}; also the chosen time-dependence is
natural for some systems typical of quantum optics under the so called
heterodyne/homodyne detection scheme.

\section{Entropy and information gain}\label{S3}

In \cite{Oza86} a measurement is called quasi-complete if the \emph{a
posteriori} states are pure for every pure initial state and it is called
complete if the \emph{a posteriori} states are pure for every (pure or mixed)
initial state. So, we call \emph{quasi-complete} the continual measurement of
Section \ref{S2} if the \emph{a posteriori} states $\rho_t$ are \emph{pure}
($P_\rho$-almost surely) \emph{for all $t$ and for all pure initial conditions}
$\rho$. In \cite{BarP00} we proved that
\begin{theorem}\label{theo1}
The continual measurement of Section \ref{S2} is quasi-complete if and only  if
$\mathcal{L}_1 = 0$ and $ \displaystyle
\frac{J_k[\rho]}{\Tr\left\{J_k[\rho]\right\}}$ is a pure state for every $k$
and for every pure state $\rho$. In this case there exists a partition $A_1,
A_2$ of the integer numbers such that  for some $R_k \in
\mathcal{B}(\mathcal{H})$ and for some monodimensional projection $P_k$ we can
write $ J_k[\rho] =  R_k \rho R_k^\dagger$, for $k \in A_1$, $J_k[\rho]
=\Tr\left\{ \rho J^*_k[\bbbone]\right\} P_k$, for $ k \in A_2$.
\end{theorem}

Our continual measurement can not be complete in the sense of \cite{Oza86} for
a fixed time; however, it can be ``asymptotically complete". Examples of this
behaviour in the case of linear systems are given in \cite{DohTPW99}. In
\cite{BarP00}, we proved that
\begin{theorem}\label{theo2}
Let the continual measurement of Section \ref{S2} be qua\-si-com\-ple\-te and
let $\mathcal{H}$ be finite-dimensional. If for every time $t$ it does not
exist a bidimensional projection $P_t$ such that, $\forall j, k$,
$
P_t\left(\widetilde L_j(t) + \widetilde L_j(t)^\dagger\right)P_t = z_j(t) P_t$,
$P_tJ^*_k[\bbbone]P_t = q_k(t) P_t$ for some complex numbers $z_j(t)$ and
$q_k(t)$, then eq.\ (\ref{2+12}) maps asymptotically, for $t \to \infty$, mixed
states into pure ones, in the sense that for every initial condition $\rho$ we
have $P_\rho$-almost surely
$
\lim_{t \to \infty} \Tr\left\{ \rho_{t}\left( \bbbone - \rho_{t}\right)
\right\} = 0$.
\end{theorem}

The proof of the theorems above is based on the study of the \emph{a posteriori
linear entropy} (or \emph{purity}) $\Tr\{\rho_t(\bbbone - \rho_t)\}$ and of its
mean value. However, physically  more interesting quantities are the von
Neumann entropy and the relative entropy: for $x,y\in
\mathcal{S}(\mathcal{H})$, $S[x]= - \Tr \left\{x \ln x \right\}\geq 0$,
$S[x|y]=  \Tr \left\{x \ln x - x \ln y \right\}\geq 0$ (they can also diverge)
\cite{AliL87}. In our case we have the initial state $\rho = \rho_0=
\sigma_0=\eta_0$ and the \emph{initial entropy} $S[\rho]$, the \emph{a priori}
state $\eta_t$ and the \emph{a priori entropy} $S[\eta_t]$, the \emph{a
posteriori} states $\rho_t$ and the \emph{mean a posteriori entropy}
\begin{equation}\label{3+4}
\mathbb{E}_{P_\rho}\big[ S[\rho_t]\big] = \mathbb{E}_Q \big[ \|\sigma_t\| \ln
\| \sigma_t\| - \Tr \{ \sigma_t \ln \sigma_t\}\big]\,.
\end{equation}
By some direct computations, we obtain a first relation among these quantities:
\begin{equation}\label{3+9}
S[\eta_t]- \mathbb{E}_{P_\rho}\big[S[\rho_t]\big] = \mathbb{E}_{P_\rho}
\big[S[\rho_t|\eta_t]\big]\geq 0\,.
\end{equation}

Following \cite{Oza86}, we can also introduce the \emph{amount of information}
of the continual measurement
\begin{equation}\label{3+5}
I[\rho;t] = S[\rho] - \mathbb{E}_{P_\rho}\big[ S[\rho_t]\big]
\end{equation}
and the classical amount of information. To introduce this last quantity we
need some notations. Let us set $P_\rho(\dd \omega ;t) = \|\sigma_t(\omega)\|
Q(\dd \omega)$, let $\rho=\sum_\alpha w_\alpha \rho_\alpha$ be the orthogonal
decomposition of $\rho$ into pure states and $P_{\rho_\alpha}$,
$\sigma_t^\alpha$, $\rho_t^\alpha$, $\eta_t^\alpha$, $m_j^\alpha(t)$,
$\nu_k^\alpha(t)$ be defined starting from $\rho_\alpha$ as $P_{\rho}$,
$\sigma_t$, $\rho_t$, $\eta_t$, $m_j(t)$, $\nu_k(t)$ are defined starting from
$\rho$. Then, the \emph{classical amount of information} of the continual
measurement is defined by
\begin{eqnarray}\nonumber
\mbox{\textrm{c-}}I[\rho;t] &=& \sum_\alpha w_\alpha \int_\Omega \ln \left(
\frac{ P_{\rho_\alpha}(\dd \omega;t)} { P_{\rho}(\dd \omega;t)}\right)
P_{\rho_\alpha}(\dd \omega;t)
\\ \nonumber
&=& \sum_\alpha w_\alpha \mathbb{E}_{P_{\rho_\alpha}} \left[ \ln \frac
{\|\sigma_t^\alpha\|} { \|\sigma_t\|} \right]
\\ \label{3+6}
&=& \mathbb{E}_Q\left[ \sum_\alpha w_\alpha \|\sigma_t^\alpha\| \ln
\|\sigma_t^\alpha\| -  \|\sigma_t\| \ln \|\sigma_t\|\right].
\end{eqnarray}

By classical arguments, $ \mbox{\textrm{c-}}I[\rho;t]$ is always positive
\cite{Oza86}:
$
\mbox{\textrm{c-}}I[\rho;t] \geq 0$, $\forall t\geq 0$, $\forall \rho\in
\mathcal{S}(\mathcal{H})$. Obviously, we have $I[\rho;t] \leq S[\rho]$,
$I[\rho;0]=0$, $\mbox{\textrm{c-}}I[\rho;0]=0$. If it exists an equilibrium
state $\eta_{\mathrm{eq}}$ ($\mathcal{L}[\eta_{\mathrm{eq}}]=0$), by
(\ref{3+9}) we have also $I[\eta_{\mathrm{eq}};t]\geq 0$.

\begin{theorem}\label{theo4}
The classical amount of information of the continual measurement of Section
\ref{S2} is non-decreasing in time and {\rm
\begin{eqnarray}\nonumber
\frac{\dd\ }{\dd t}\,\mbox{\textrm{c-}}I[\rho;t] &=& \sum_\alpha w_\alpha
\mathbb{E}_{P_{\rho_\alpha}} \left[ \frac 1 2 \sum_j m_j^\alpha(t)^2 + \sum_k
\nu_k^\alpha (t) \ln \nu_k^\alpha (t) \right]
\\ \nonumber
{}&-& \mathbb{E}_{P_{\rho}} \left[ \frac 1 2 \sum_j m_j(t)^2 + \sum_k \nu_k (t)
\ln \nu_k (t) \right]
\\ \nonumber
&=& \sum_\alpha w_\alpha \mathbb{E}_{P_{\rho_\alpha}} \left[ \frac 1 2 \sum_j
\left( m_j^\alpha(t)-m_j(t)\right)^2\right.
\\ \label{3.10}
{}&+& \left. \sum_k \nu_k(t) \left( 1 - \frac{\nu_k^\alpha (t)}{\nu_k(t)} +
\frac{\nu_k^\alpha (t)}{\nu_k(t)} \ln \frac{\nu_k^\alpha (t)}{\nu_k(t)}\right)
\right] \geq 0\,.
\end{eqnarray}}
\end{theorem}
To prove this theorem one has to differentiate the last expression in
(\ref{3+6}) and to use the relationships among $Q$, $P_\rho$,
$P_{\rho_\alpha}$.

For quasi-complete measurements the information gain $I[\rho;t]$ has a nice
behaviour.
\begin{theorem}\label{theo3}
The continual measurement of Section \ref{S2} is quasi-complete if and only if
the amount of information $I[\rho;t]$ is non-negative for any $\rho\in
\mathcal{S}(\mathcal{H})$ with $S[\rho]< +\infty$ and any $t\geq 0$. Moreover,
if it is quasi-complete, we have $I[\rho;t]\geq \mbox{\rm c-}I[\rho;t] \geq 0$,
$I[\rho;t]\geq I[\rho;s]$ for any t, any $s<t$ and any state $\rho$  with
$S[\rho]< +\infty$.
\end{theorem}
\textbf{Proof.} All the statements but the last one are a particularization of
Theorems 1 and 2 of \cite{Oza86} to our case. The last statement needs the use
of conditional expectations. We have $I[\rho;t]- I[\rho;s]=
\mathbb{E}_{P_\rho}\big[ S[\rho_s]
-\mathbb{E}_{P_\rho}[S[\rho_t]|\mathcal{F}_s]\big]$; by (\ref{2+12}) $S[\rho_s]
-\mathbb{E}_{P_\rho}[S[\rho_t]|\mathcal{F}_s]$ is the amount of information at
time $t$ when the initial time is $s$ and the initial state is $\rho_s$ and,
so, it is non-negative for a quasi-complete measurement. \hfill{$\square$}

\smallskip

Finally, if $\mathcal{H}$ is finite-dimensional, the vanishing of the purity
implies the vanishing of the entropy; therefore, we have the asymptotic
completeness also in the sense of the vanishing of the entropy:
\begin{itemize}
\item[] \emph{The hypotheses of Theorem \ref{theo2} imply also that
$\lim_{t\to+\infty} S[\rho_t] = 0$, $P_\rho$-almost surely, and
$\lim_{t\to+\infty} I[\rho;t]= S[\rho]$.}
\end{itemize}

\end{document}